\documentclass[12pt]{JHEP3}
\usepackage{graphicx}%
\usepackage{epsfig}
\epsfclipon
\usepackage{multicol}
\usepackage{epsfig,bm}
\usepackage{amssymb,amsmath}
\newcommand{\roughly}[1]{\mathrel{\raise.3ex\hbox{$#1$\kern-0.85em
\lower1ex\hbox{$\sim$}}}}

\newcommand{\bmat}{\left(\begin{array}}
\newcommand{\emat}{\end{array}\right)}

\def\yzero{\smash{\hbox{$y\kern-4pt\raise1pt\hbox{${}^\circ$}$}}}

\def\beq{\begin{equation}}
\def\eeq{\end{equation}}
\def\beqa{\begin{eqnarray}}
\def\eeqa{\end{eqnarray}}

\def\-{\hphantom{-}}

\def\s2{\frac{1}{2}}

\def\IF{\relax{\rm I\kern-.18em F}}
\def\II{\relax{\rm I\kern-.18em I}}
\def\IP{\relax{\rm I\kern-.18em P}}
\def\IC{\relax{\rm I\kern-.48em C}}
\def\IR{\relax{\rm I\kern-.18em R}}

\def\cn{{\cal N}}

\def\cp{{\cal P}}

\def\Dsl{\,\raise.15ex\hbox{/}\mkern-13.5mu D} 
\def \one{\relax{\rm 1\kern-.26em I}}

\def\pref#1{(\ref{#1})}

 \def\cp#1{\relax\ifmmode {\IP\kern-2pt{}_{#1}}\else $\IP\kern-2pt{}_{#1}$\=fi}

\catcode`\@=11
\newdimen\@rotdimen
\newbox\@rotbox

\def\@vspec#1{\special{ps:#1}}
\def\@rotstart#1{\@vspec{gsave currentpoint currentpoint translate
   #1 neg exch neg exch translate}}
\def\@rotfinish{\@vspec{currentpoint grestore moveto}}
\def\@rotr#1{\@rotdimen=\ht#1\advance\@rotdimen by\dp#1%
   \hbox to\@rotdimen{\hskip\ht#1\vbox to\wd#1{\@rotstart{90 rotate}%
   \box#1\vss}\hss}\@rotfinish}
\def\@rotl#1{\@rotdimen=\ht#1\advance\@rotdimen by\dp#1%
   \hbox to\@rotdimen{\vbox to\wd#1{\vskip\wd#1\@rotstart{270 rotate}%
   \box#1\vss}\hss}\@rotfinish}%
\def\@rotu#1{\@rotdimen=\ht#1\advance\@rotdimen by\dp#1%
   \hbox to\wd#1{\hskip\wd#1\vbox to\@rotdimen{\vskip\@rotdimen
   \@rotstart{-1 dup scale}\box#1\vss}\hss}\@rotfinish}%
\def\@rotf#1{\hbox to\wd#1{\hskip\wd#1\@rotstart{-1 1 scale}%
   \box#1\hss}\@rotfinish}%
\def\rotate{\@ifnextchar[{\@rotate}{\@rotate[l]}}
\def\@rotate[#1]#2{\setbox\@rotbox=\hbox{#2}\@nameuse{@rot#1}\@rotbox}

\catcode`\@=12

\title{Racetrack Inflation}

\author{J.J.~ Blanco-Pillado,$^1$ C.P.~Burgess,$^2$ J.M.~Cline,$^2$
C.~Escoda,$^3$ M.~G\'omez-Reino,$^4$ R.~Kallosh,$^5$ A.~Linde,$^5$
and F.~Quevedo$^3$
\\

$^1$ Center for Cosmology and Particle Physics,
New York University, New York, NY 10003, USA\\

$^2$Physics Department, McGill University, 3600 University Street,
Montr{\'e}al, Qu{\'e}bec, H3A 2T8, Canada \\

$^3$ DAMTP, Centre for Mathematical Sciences,
University of Cambridge, Cambridge CB3 0WA, UK\\

$^4$ Martin Fisher School of Physics,
Brandeis University, Waltham MA 02454, USA\\

$^5$ Department of Physics, Stanford University, Stanford, CA 94305--4060, USA}

\abstract{We develop a model of eternal topological inflation
using a racetrack potential within the context of type IIB string
theory with KKLT volume stabilization. The inflaton field is the
imaginary part of the K\"ahler structure modulus, which is an
axion-like field in the 4D effective field theory. This model does
not require moving branes, and in this sense it is simpler than
other models of string theory inflation. Contrary to
single-exponential models, the structure of the potential in this
example allows for the existence of saddle points between two
degenerate local minima for which the slow-roll conditions can be
satisfied in a particular range of parameter space. We conjecture
that this type of inflation should be present in more general
realizations of the modular landscape. We also consider
`irrational' models having a dense set of minima, and discuss
their possible relevance for the cosmological constant problem.}

\preprint{BRX TH-544, DAMTP-2004-68, McGill-04/15\\  hep-th/0406230}

\begin{document}

\makeatletter \@addtoreset{equation}{section} \makeatother
\renewcommand{\theequation}{\thesection.\arabic{equation}}







\setcounter{page}{1} \pagestyle{plain}
\renewcommand{\thefootnote}{\arabic{footnote}}
\setcounter{footnote}{0}


\section{Introduction}

The past three years have seen a revival of attempts to derive
cosmological inflation from string theory. On the theoretical
side, the main reason for this effort has been the development of
tools for studying modulus stabilization, through the explicit
construction of nontrivial potentials for moduli fields within
string theory. From the observational side, the impetus has been
the successful inflationary description of recent CMB results.

Early attempts to find inflation within string theory looked to
the string dilaton $S$ and the geometrical moduli fields (such as
the volume modulus, $T$) as natural candidates for the inflaton
field \cite{bg,banks}. However these attempts were largely
thwarted by the flatness (to all orders in perturbation theory) of
the corresponding scalar potentials, together with the discovery
that the few calculable nonperturbative potentials considered
\cite{drsw}-\cite{filq1} were not flat enough to satisfy the
slow-roll conditions needed for inflation. This led to the
alternative proposal of D-brane inflation, which instead considered the
separation between branes as an inflaton candidate
\cite{dvalitye}-\cite{dbraneinflation}.

In the absence of an understanding of how moduli are fixed, the
main working assumption used when exploring these proposals was
simply that all moduli aside from the putative inflaton were
fixed by an unknown mechanism at large scales compared with those
relevant to inflation. In particular it was assumed that this
fixing could be ignored when analyzing the inflaton dynamics. It
is now possible to do better than this, following the
recent study of brane/antibrane inflation \cite{kklmmt} (see also
\cite{renata}-\cite{shamit}) which could follow the interplay
between the inflaton and other moduli by working within the KKLT
scenario for moduli fixing \cite{kklt}.

In so doing these authors discovered an obstacle to the successful
realization of inflation, which is a version of the well-known
$\eta$ problem of $F$-term inflation within supersymmetric theories
\cite{etaproblem}. The problem arises because the structure of the
supergravity potential (see {\it e.g.} eq.~\pref{treepot}) involves
an overall factor proportional to $e^K$ where $K$ is the K\"ahler
potential, and this naturally induces a mass term which is of order
the Hubble scale, $H$, for the inflaton field. As such it
contributes a factor of order one to the slow-roll parameter
$\eta$, which must be very small in order for the model to agree with observations.
Consequently, a
fine tuning of roughly 1 part in 100 in \cite{kklmmt} and in 1 part in 1000 in \cite{bcsq} is required in
these models for a successful inflation. This tuning
would be unnecessary if the inflaton field did not appear within
the K\"ahler potential. Unfortunately $K$ {does} depend on the
inflaton in the D3/anti-D3 systems considered in \cite{kklmmt,bcsq}
due to the absence of isometries in Calabi-Yau spaces. A brane
position can avoid appearing within the K\"ahler potential as well as non-perturbative superpotential
  for the
system of D3/D7 branes \cite{d3d7}. This is  due to the underlying ${\cal N}=2$
supersymmetry  of these systems, and therefore for these model there is
no $\eta$ problem  \cite{renata}.

In this article we present a different approach for string
inflation for which a geometrical modulus is the inflaton, without
the need of introducing the interacting D-branes and their
separation. In this way we revive the original proposals for
modular inflation of \cite{bg,banks} (see also \cite{bmqrz}). Our
proposal is based on a simple extension of the KKLT scenario to
include a racetrack-type superpotential, along the lines developed
in \cite{egq}. The difficulty with obtaining inflation using the
simplest potentials considered in the past \cite{drsw}-\cite{filq1} --- as well as more recently in \cite{kklt} ---
is that the potential is never flat enough to allow for slow roll.
However in nonperturbative potentials of the modified
racetrack\footnote{By `modified racetrack' we mean a racetrack
superpotential \cite{racetrack} -- {\it i.e.} containing more than one exponential
of the K\"ahler modulus -- which is modified by adding a constant
term \cite{kklt,egq}, such as arises from the three-form fluxes of the type IIB
compactifications.} type  such as arise in type
IIB string compactifications we find saddle points which give rise
precisely to the conditions for topological inflation
\cite{andrei,vilenkin},  being flat enough to provide the right
number of $e$-foldings and the flat spectrum of the perturbations of the metric.
An attractive feature of using a modulus as inflaton is that the
tree-level K\"ahler potential can be independent of some fields,
like the imaginary part of the K\"ahler modulus field Im $T$, and
so at this level the $\eta$ problem is not necessarily present.
(However approximate symmetries such as shifts in Im $T$ are
typically broken both by the nonperturbative superpotential and
by loop corrections to the K\"ahler potential.) We are led in this
way to an inflaton field which is an axion-like pseudo-Goldstone
mode. In this respect, our scenario resembles the natural
inflation scenario \cite{nat}.

A feature of natural inflation scenarios which is {\it not} shared
by the inflation which we find is the assumption that only the
axion field evolves, with the pseudo-Goldstone mode running along
the valley of the potential for which all other fields are
stabilized. We find no such regime so far in supergravity or
string theory, and instead in our scenario we find that the volume
modulus, Re $T$, is stabilized only in the vicinity of the KKLT
minima and near a saddle point of the potential in between these
minima, where the axion field vanishes. Inflation does not occur
near the KKLT minima but, as we shall see, under certain
conditions it can occur near the saddle points.

We start, in section 2, with a short recapitulation of topological
inflation \cite{andrei,vilenkin} before discussing our main
results about racetrack inflation in section 3. In section 4 we
present an interesting generalization of our results to include an
`irrational' dependence on the inflaton field leading to an
infinite number of vacua, and we discuss there its possible
implications for our inflationary scenario as well as for the
cosmological constant problem.

\section{Eternal Topological Inflation}\label{topinfl}

Slow-roll inflation is realized if a scalar potential, $V(\phi)$,
is positive in a region where the following conditions are
satisfied:
\begin{equation}
\label{slowroll}
\epsilon  \equiv  \frac{M_{pl}^2}{2}\ \left( \frac{V'}{V}\right)^2\
\ll  \ 1\ , \ \ \ \  \qquad
\eta  \equiv  M_{pl}^2\ \frac{V''}{V}\  \ll  \ 1\ .
\end{equation}
Here $M_{pl}$ is the rationalized Planck mass ($(8\pi G)^{-1/2}$)
and primes refer to
derivatives with respect to the scalar field, which is assumed to
be canonically normalized. Satisfying these conditions is not an
easy challenge for typical potentials since the inflationary
region has to be very flat. Furthermore, after finding such a
region we are usually faced with the issue of initial conditions:
Why should the field $\phi$ start in the particular slow-roll
domain?

For the simplest chaotic inflation models of the type of ${m^2\over 2} \phi^2$
this problem can be easily resolved. In these theories, inflation may start  if
one has a single domain at a nearly Planckian density, where the field is large
and homogeneous on a scale that can be as small as the Planck scale. One can
argue that the probability of this event should not be strongly suppressed
\cite{initial,book}. Once inflation begins, the universe enters the process of
eternal self-reproduction due to quantum fluctuations which unceasingly return
some parts of the universe to the inflationary regime. The total volume of
space produced by this process is infinitely greater than the total volume of
all non-inflationary domains \cite{Eternal,LLM}.

The problem of initial conditions in the theories where inflation is possible
only at the densities much smaller than the Planck density is much more
complicated. However, one may still argue that even if the probability of
proper initial conditions for inflation is strongly suppressed, the possibility
to have eternal inflation infinitely rewards those domains where inflation
occurs. One may argue that the problem of initial conditions in the theories
where eternal inflation is possible is largely irrelevant, see \cite{LLM} for a
discussion of this issue.

Eternal inflation is not an automatic property of all inflationary
models. Many versions of the hybrid inflation scenario, including
some of the versions used recently for the implementation of
inflation in string theory, do not have this important property.
Fortunately,  inflation is eternal in all models where it occurs
near the flat top of an effective potential. Moreover, it occurs
not only due to quantum fluctuations \cite{VilEternal}, but even
at the classical level, due to eternal expansion of topological
defects \cite{andrei,vilenkin}.

The essence of this effect is very simple. Suppose that the
potential has a saddle point, so that some components of the field
have a small tachyonic mass $|m^2| \ll H^2$, where $H^2 = {V/( 3
M_{pl}^2)}$, and $V$ is the value of the effective potential at
the saddle point. Consider for example a sinusoidal wave of the
field $\phi$, $\delta\phi \sim \phi_0 \sin kx$ with $k \ll m$. It
can be shown that the amplitude of such a wave grows as $e^{|m^2|t /
3H}= e^{\eta H t}$ \cite{book}, whereas the distance between the nodes of this
wave grows much faster, as $e^{Ht}$. As a result, at each
particular point the field falls down and inflation ends, but the
total volume of all points staying close to the saddle point
continues growing exponentially, making inflation eternal
\cite{andrei,vilenkin}.

\section{Racetrack Inflation}

In this section we exhibit an example of topological inflation
within string moduli space, following the KKLT scenario.

\subsection{The Effective 4D Theory}

Recall that this scenario builds on the GKP construction
\cite{gkp} (see \cite{sethi1} for earlier discussions), for which
type IIB string theory is compactified on an orientifolded
Calabi-Yau manifold in the presence of three-form RR and NS
fluxes, as well as D7 branes containing $\cn=1$ supersymmetric
gauge field theories within their world-volumes. The background
fluxes provide potential energies which can fix the values of the
complex dilaton field and of the complex-structure moduli. The
resulting effective 4D description of the K\"ahler moduli is a
supergravity with a potential of the no-scale type \cite{noscale},
corresponding to classically flat directions along which
supersymmetry generically breaks. KKLT start with Calabi-Yaus
having only a single K\"ahler modulus, and lift this remaining
flat direction using nonperturbative effects to induce nontrivial
superpotentials for it. After fixing this modulus they introduce
anti-D3 branes to lift the minimum of the potential to nonnegative
values, leading to a metastable de Sitter space in four
dimensions. Alternatively, the same effect as the anti-D3 branes
can be obtained by turning on magnetic fluxes on the D7 branes,
which give rise to a Fayet-Iliopoulos D-term potential \cite{bkq}.
Other combinations of non-perturbative effects in string theory leading to dS vacua were proposed in \cite{Brustein:2004xn}.

We here follow an identical procedure, based on the dynamics of a
single K\"ahler modulus, $T$, whose real part measures the volume
of the underlying Calabi-Yau space.\footnote{It has been recently
argued \cite{douglas,sethi} that nonperturbative superpotentials
cannot be generated for a large class of one-modulus Calabi-Yau
compactifications, with the authors of these references differing
on whether or not the resulting landscape is half-full or
half-empty. We do not regard their results as an air-tight no-go
theorem for single-modulus vacua until more exhaustive studies of
string vacua are performed. For instance, mechanisms like
orbifolding and turning on magnetic fluxes on D7 branes could
modify the matter spectrum of the $N=1$ supersymmetric theory
within the D7 brane in such a way that nonperturbative
superpotentials of the gaugino condensation type could be induced.
These issues are now investigated in \cite{GKTT}.
In any case, since the single-modulus examples are the simplest
scenarios, they can always be seen as some sort of limiting
low-energy region ---  the one for which all but one of the
K\"ahler moduli have been fixed at higher scales --- in the
many-moduli compactifications that have been found to lead to
nonperturbative superpotentials.} (For type IIB theories the
imaginary part of $T$ consists of a component of the RR 4-form
which couples to 3-branes.) Just as for KKLT, all other fields are
assumed to have been fixed by the background fluxes, and a
superpotential, $W(T)$, for $T$ is imagined to be generated, such
as through gaugino condensation \cite{drsw,ourgc} within the gauge
theories on the D7 branes (which depend on $T$ because the
volume, Re~$T$, defines the gauge coupling on the D7 branes).

Our treatment differs from KKLT only in the form assumed for the
nonperturbative superpotential, which we take to have the
modified racetrack form \cite{racetrack,egq}
\begin{equation}\label{dosexp}
W=W_0+A\,e^{-aT}+B\,e^{-bT}\,.
\end{equation}
such as would be obtained through gaugino condensation in a theory
with a product gauge group. For instance, for an $SU(N)\times
SU(M)$ group we would have $a=2\pi/N$ and $b=2\pi /M$. Because the
scale of $A$ and $B$ is set by the cutoff of the effective theory,
we expect both to be small when expressed in Planck units
\cite{gauginoproduct}. The constant term $W_0$ represents the
effective superpotential as a function of all the fields that have
been fixed already, such as the dilaton and complex structure
moduli.

This superpotential includes the one used by KKLT as the special
case $AB=0$. It also includes the standard racetrack scenario
(when $W_0=0$), which was much discussed in order to fix the
dilaton field at weak coupling in the heterotic string
\cite{racetrack}. Its utility in this regard is seen for large
values of $N$ and $M$, with $M$ close to $N$, since then the
globally-supersymmetric minimum, $W' = 0$, occurs when
\begin{equation}
T\ =\ \frac{NM}{M-N} \log \left(-\frac{MB}{NA}\right)
\end{equation}
and so is guaranteed to lie in the region where Re~$T$ is large,
corresponding to weak coupling.\footnote{This is the main idea
behind the standard racetrack scenarios, originally proposed for
the heterotic string in order to get minima with small gauge
coupling constant. Notice that the large value of $T$
corresponding to $W'=0$ is independent of the value of $W_0$,
which was not introduced in the original racetrack models but
plays an important role here.} The same proves to be true for
minima of the full supergravity potential, for which simple
analytic expressions are not available. This superpotential was
first adapted to the KKLT scenario in \cite{egq}.

Following KKLT, the scalar potential we consider is a sum of two
parts
\begin{equation}
V= V_F + \delta V \,.
\end{equation}
The first term comes from the standard $\cn=1$ supergravity
formula for the F-term potential, which in Planck units \cite{Cremmer}
reads
\begin{equation}
V_{F}~=~e^{K}\left( \sum_{i,j} K^{i\bar{j}} D_{i}W \overline {D_j W}
- 3|W|^2 \right)\,,
\label{treepot}
\end{equation}
where $i,j$ runs over all moduli fields, $K$ is the K\"ahler
potential for $T$, $K^{i\bar{j}}$ is the inverse of
$\partial_i\partial_{\bar{j}}K$, and
$D_iW=\partial_iW+(\partial_iK)W$. For the K\"ahler potential,
$K$, we take the weak-coupling result obtained from Calabi-Yau
compactifications \cite{kahlerform}, namely
\begin{equation}
K=-3\log (T+T^*)\,. \label{kpdef}
\end{equation}
We neglect the various possible perturbative and nonperturbative
corrections to this form which might arise.

The nonsupersymmetric potential, $\delta V$, is that part of the
potential which is induced by the tension of the anti-D3
branes.\footnote{In \cite{bkq} the anti-D3 brane was substituted
by magnetic fluxes on D7 branes. The potential generated is
identical to the one induced by the anti-D3 brane, with the
advantage of having the interpretation of a supersymmetric
Fayet-Iliopoulos D-term.} The introduction of the anti-brane does
not introduce extra translational moduli because its position is
fixed by the fluxes \cite{kpv}, so it just contributes to the
energy density of the system. This contribution is positive
definite and depends on a negative power of the Calabi-Yau volume,
$X = \hbox{Re}\,T$, as follows \cite{kpv}:
\begin{equation}
\delta V = { E\over X^\alpha}\label{sb}\,,
\end{equation}
where the coefficient $E$ is a function of the tension of the
brane $T_3$ and of the warp factor. The exponent $\alpha$ is
either $\alpha=2$ if the anti-D3 branes are sitting at the end of
the Calabi-Yau throat, or in the case of magnetic field fluxes
\cite{bkq}, if the D7 branes are located at the tip of the throat.
Otherwise $\alpha=3$ corresponding to the unwarped region (in the
anti-D3 brane case the warped region is energetically preferred
and we will usually take $\alpha=2$). There is clearly
considerable model dependence in this scenario. It depends on the
number of K\"ahler moduli of the original Calabi-Yau manifold, on
what kind of nonperturbative superpotential can be induced for
them (if any), on the K\"ahler function, the power $\alpha$, and so
on.

\subsection{The Scalar Potential}

We now explore the shape of the scalar potential, to identify
potential areas for slow-roll inflation. To this end we write the
field $T$ in terms of its real and imaginary parts:
\begin{equation}
T\, \equiv X+i Y\,.
\end{equation}
Notice that, to the order that we are working, the K\"ahler
potential depends only on $X$ and not on $Y$. For fields rolling
slowly in the $Y$ direction this feature helps address the $\eta$
problem of F-term inflation.

Using (\ref{treepot}) and (\ref{kpdef}) the scalar potential turns
out to be
\begin{equation}
V_{F}  =  \frac1{8 X^3} \left\{ \frac13|2X W^\prime  - 3 W|^2 - 3|W|^2
\right \}\,,
\label{spot}\end{equation}
where $^\prime$ denotes derivatives with respect to $T$. The
supersymmetric configurations are given by the solutions to
\begin{equation}
2X W^\prime - 3 W = 0\,.
\label{msloc}
\end{equation}
Using only $V_F$, the values of the potential at these
configurations are either negative or zero, corresponding to
anti-de Sitter or Minkowski vacua. Substituting the explicit form
of the superpotential  and adding the SUSY-breaking term
we find that the scalar potential becomes
\begin{eqnarray}
    V~&= & \frac{E}{X^\alpha}\  +\
    \frac{e^{-aX}}{6X^2}\left [ aA^2\left(aX+3\right)~ e^{-aX}\ +
    3 W_0 aA \cos(aY)\right] + \nonumber \\
    &+&\frac{e^{-bX}}{6X^2}\left [ bB^2\left(bX+3\right)~ e^{-bX}\
    + 3 W_0 bB \cos (bY)\right] + \nonumber \\
    &+&\frac{e^{-(a+b)X}}{6X^2}\left[ AB\left(2abX +3a +3b\right)
    \cos((a-b)Y) \right]
\label{potential1}
\end{eqnarray}

This potential has several de Sitter (or anti-de Sitter) minima,
depending on the values of the parameters $A,a,B,b,W_0, E$. In
general it has a very rich structure, due in part to the
competition of the different periodicities of the $Y$-dependent
terms. In particular, $a-b$ can be very small, as in standard
racetrack models, since we can choose $a=2\pi /M$, $b=2\pi/N$ with
$N\sim M$ and both large integers. Notice that in the limit
$(a-b)\rightarrow 0$ and $W_0 \rightarrow 0$, the $Y$ direction
becomes exactly flat. We can then tune these parameters (and $AB$)
in order to obtain flat regions suitable for inflation.

From the above we expect extrema situated at large
$X=$ Re~$T$ given
a discrete fine tuning of $M$ and $N$. This is independent of the
value of $W_0$ and in particular occurs for $W_0=0$. This
behaviour is different from the original KKLT scenario, which does
not have any minima when $W_0=0$. For $W_0\neq 0$ many new local
minima appear due to the periodicity of the terms proportional to
$W_0$ in the scalar potential. For the minima of interest we use
the freedom to choose the value of $E$, as in KKLT, to tune the
global minimum of the potential to the present-day vacuum energy.

Furthermore, the shape of the potential is very sensitive to the values of
the parameters. We find that the potential has a maximum in the $Y$-
direction
if  the following conditions are satisfied:
$A+B<0$, $W_0<0$, $a<b$. Otherwise the point $Y=0$ would correspond to
a minimum. With these condtions satisfied and for a fixed value of
the other parameters, there is a critical value  of $W_0$ beyond which
 the
point $Y=0$ is also a maximum in the $X$-direction and therefore the
field runs away to the uncompactified limit $X\rightarrow \infty$.
But for $W_0$ smaller than the critical value, the point $Y=0$ is a
minimum in the $X$-direction and therefore a saddle point. This is the
interesting range to look for slow-roll inflation.

\begin{figure}[h!]
\centering\leavevmode\epsfysize=10cm \epsfbox{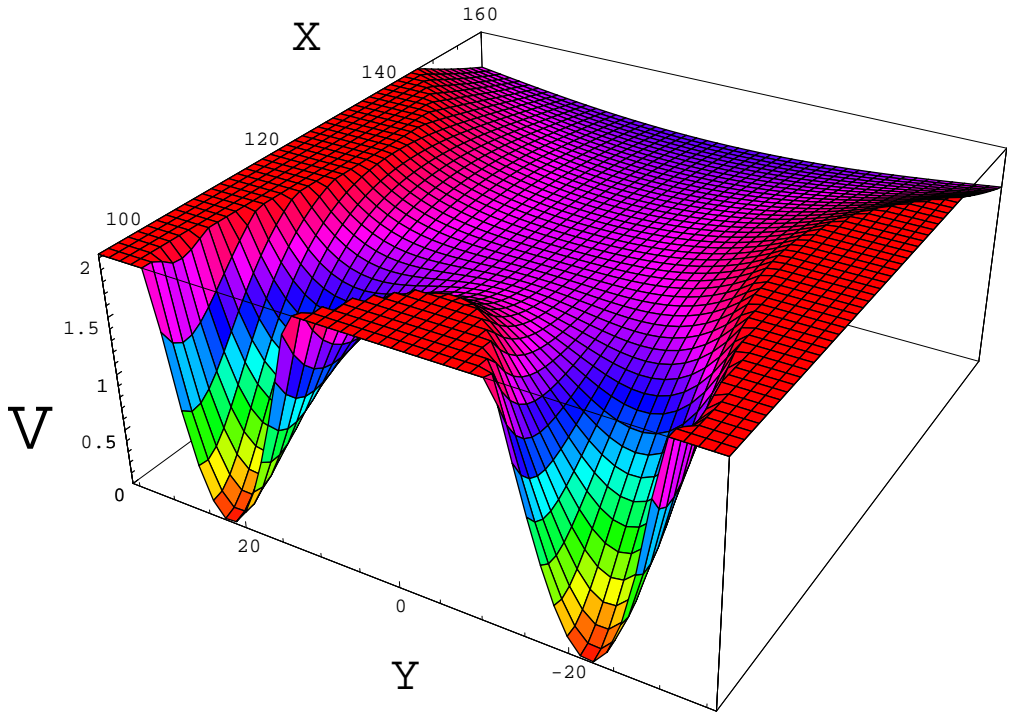}

\caption[fig1] {Plot for a racetrack type potential (rescaled by
$10^{16}$). Inflation begins in a vicinity of the saddle point at $X_{\rm saddle}=123.22$, $ Y_{\rm saddle}=0$.
Units are $M_p=1$.\label{F1}}
\end{figure}

\begin{figure}[h!]
\centering\leavevmode\epsfysize=6cm \epsfbox{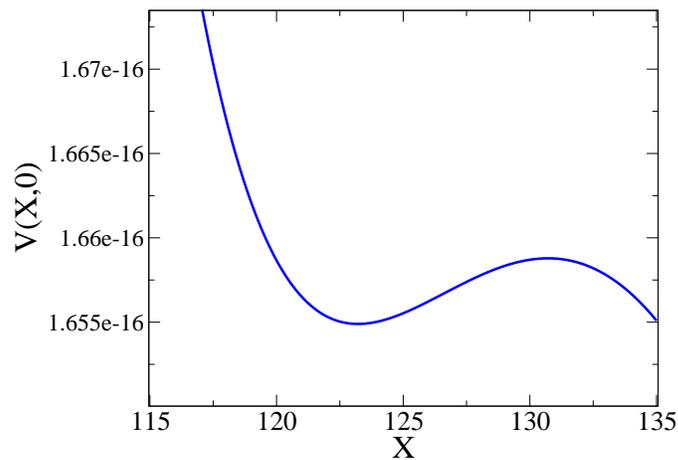}

\caption[fig2] {Plot of $V(X,0)$ versus $X$ in the vicinity of the saddle point $X_{\rm saddle}=123.22$, $ Y_{\rm saddle}=0$.\label{F2}}
\end{figure}

Figs.~\ref{F1}-\ref{F2} illustrate a region of the scalar potential for
which inflation is possible. The values of the parameters which
are used to obtain this potential are:
\begin{equation}
\label{parvalues}
A=\frac{1}{50}, \qquad B=-\, \frac{35}{1000}, \qquad
a=\frac{2\pi}{100}, \qquad b=\frac{2\pi}{90},
  \qquad W_0 = -\frac{1}{25000}\   \ .
\end{equation}
With these values the two minima seen in the figure occur for
field values
\begin{equation}
X_{min}\ =\ 96.130, \qquad\quad Y_{min}\ =\ \pm 22.146 \ ,
\end{equation}
and the inflationary saddle point is at 
\begin{equation}
\label{saddlepoint}
    X_{\rm saddle}=123.22, \qquad \quad Y_{\rm saddle}=0 \,.
\end{equation}

The value of $E$ is fixed by demanding that the value of the
potential at this minimum be at zero, which turns out to require
$E = 4.14668 \times 10^{-12}$. We find this to be a reasonable value
given that $E$ is typically suppressed by the warp factor of the
metric at the position of the anti-brane.

It is crucial that this model contains two degenerate minima\footnote{Note that
the potential  is periodic with period 900, i. e. there is a set of two
degenerate minima at every  $Y=900\; n$ where $n=0, 1, 2, ...$. etc.}  since
this guarantees the existence of causally disconnected regions of space which
are in different vacua.  These regions necessarily have a domain wall between
them where the field is near the saddle point and thus eternal inflation is
taking place, provided that the slow roll conditions (\ref{slowroll}) are
satisfied there.  It is then inevitable to have regions close to the saddle in
which inflation occurs, with a sufficiently large duration to explain our flat
and homogeneous universe.

\subsection{Scaling Properties of the Model}

It is easy to see from the potential (\ref{potential1}) that we can obtain
models with rescaled values of the critical points for other choices  of
parameters, having the same features of inflation. This can be done simply by
rescaling their values in the following way,
\begin{equation}
\label{rescaling1}
a \rightarrow a/\lambda \,,\hspace{.5cm}
b\rightarrow b/\lambda \,,\hspace{.5cm}
E\rightarrow \lambda^2 E \,,\hspace{.5cm}
\end{equation}
and also
\begin{equation}
\label{rescaling2}
A \rightarrow \lambda^{3/2} A \,,\hspace{.5cm}
B\rightarrow \lambda^{3/2} B \,,\hspace{.5cm}
W_0\rightarrow \lambda^{3/2} W_0 \,,\hspace{.5cm}
\end{equation}
Under all these rescalings the potential does not change under condition that
the  fields also rescale
\begin{equation}
X \rightarrow \lambda X \,,\hspace{.5cm}
Y \rightarrow \lambda Y \,,\hspace{.5cm}
\end{equation}
in which case  the location of the extrema also rescale. One can verify that
the values  of the slow-roll parameters $\epsilon$ and $\eta$ do not change and
also the amplitude of the density perturbations ${\delta \rho\over \rho}$
remains the same. It is important to take into account that the kinetic term in
this model is invariant under the rescaling, which is not  the case for
canonically normalized fields.

Another property of this model is given by the following rescalings
\begin{equation}
\label{rescaling3}
a \rightarrow a/\mu \,,\hspace{.5cm}
b\rightarrow b/\mu \,,\hspace{.5cm}
E\rightarrow  E/\mu \,,\hspace{.5cm}
\end{equation}
The potential  and the fields also rescale
\begin{equation}
\label{rescaling4}
V\rightarrow  \mu^{-3} V\,,\hspace{.5cm} X \rightarrow \mu X \,,\hspace{.5cm}
Y \rightarrow \mu Y \,,\hspace{.5cm}
\end{equation}
Under these rescalings the values of the slow-roll parameters $\epsilon$ and
$\eta$ do not change however,  the amplitude of the density perturbations
${\delta \rho\over \rho}$ scales as $\mu^{-3/2}$.

These two types of rescalings allow  to generate many other models from the
known ones, in particular, change the positions of the minima or, if one is
interested in eternal inflation, one can easily change ${\delta \rho\over
\rho}$ keeping the potential flat.

\subsection{Slow-Roll Inflation}

We now display the slow-roll inflation, by examining field motion
near the saddle point which occurs between the two minima
identified above. Near the saddle point 
$   X_{\rm saddle}=123.22, \qquad \quad Y_{\rm saddle}=0,$
the potential takes the value $V_{\rm saddle} = 1.655
\times 10^{-16}$. At this saddle point the potential has a maximum
in the $Y$ direction and a minimum in the $X$ direction, so the
initial motion of a slowly-rolling scalar field is in
the $Y$ direction.

We compute the slow-roll parameters near the top of the saddle,
keeping in mind the fact that $Y$ does not have a canonical
kinetic term. Given the K\"ahler potential we find that the kinetic
term for $X$ and $Y$ is
\beq
\label{kinterm}
{\cal L}_{\rm kin} =  {3 M_p^2\over 4 X^2}\left(\partial_\mu X \partial^\mu X
+ \partial_\mu Y \partial^\mu Y\right)
\eeq
and so
the correctly normalized $\eta$ parameter is given by the
expression $\eta = 2X^2 V''/3V$, with $X$ evaluated at the saddle
point. We find in this way the slow-roll parameters
\begin{equation}
    \epsilon_{\rm saddle}=0, \qquad \quad \eta_{\rm saddle}= -0.006097 \,,
\end{equation}
given the values of the parameters taken above. The small size of
$\eta$ is very encouraging given the reasonable range of
parameters chosen. Recall from section 2 that the slow-roll
condition $\eta \ll 1$
being satisfied at the saddle point implies automatically that we
have an inflationary regime in its vicinity, and that inflation
starting here is an example of eternal topological inflation.

\DOUBLETABLE[]{
\begin{tabular}{|c|c|c|c|}
\hline $10^3\,A$ &
$-10^3\,B$ & $-{10^{-3}\over W_{{\rm min}_{\phantom X\!\!\!}}}$ & $-{10^{-3}\over W_{\rm max}}$ \\
\hline
$20$ & $35$  & $24.998$ & $25.000$    \\
$20$ & $34$  & $20.389$ & $20.400$   \\
$10$ & $16$  & $26.766$ & $26.780$ \\
$5$  & $7.5$ & $34.496$ & $34.520$ \\
$5$  & $7$   & $21.800$ & $21.824$  \\
$3\frac13$ & $4\frac13$ & $20.280$ & $20.304$ \\
$2\frac67$ & $3\frac37$ & $14.420$ & $14.448$ \\
$2\frac67$ & $3\frac17$ & $8.6884$  & $8.708$\\
\hline
\end{tabular}}
{\begin{tabular}{|c|c|c|}
\hline
$-B$ & $-1/W_{\rm min}$ & $-1/W_{\rm max}$ \\
\hline
$3.5   $&$90685.8    $&$90667.55$\\
$3.3   $&$57273.9    $&$57286.2$ \\
$3.0   $&$27371.1    $&$27377.8$\\
$2.8   $&$16111.8    $&$16116.17$\\
$2.4   $&$5009.38    $&$5011.125$\\
$2.0   $&$1301.6     $&$1302.23$\\
$1.6   $&$267.445    $&$267.651$\\
\hline
\end{tabular}}
{A range of parameters chosen to satisfy $-0.05<\eta<0$, close to the
parameter region (3.11).\label{finetunetable1}}
{Another such range of parameters, with $a=\pi/5$, $b=2\pi/9$ and $A=1$.
\label{finetunetable2}}

How fine-tuned is this parameter choice we have found which
accomplishes inflation? Its success relies on the second
derivative, $\partial^2 V/\partial Y^2$, passing through a zero
very close to the saddle point (where $\partial V/\partial Y$
vanishes). Our ability to accomplish this relies on the freedom to
adjust $W_0$, a parameter which is not present in a pure racetrack
model. To study the range of parameters which produce inflation we
explored the vicinity of the parameters (\ref{parvalues})
%
%
which produce an inflationary solution. We were able to preserve the condition $-0.05<
\eta < 0$ by varying the parameters $W_0$ and $B$, while at all times adjusting $E$ to
keep the potential's minimum at zero. For many choices of $B$, sufficiently small
$\eta$ was obtained for $W_{\rm min} < W_0 < W_{\rm max}$, given in Table
\ref{finetunetable1}.  The same is done in Table \ref{finetunetable2} for a different
inflationary region of parameter space. We see from these tables that the success of
slow-roll inflation typically requires a fine tuning of parameters at the level of 1
part in 1000.  Parameter values in Table \ref{finetunetable1} were chosen so as to respect
the COBE normalization (see below).  This constraint was not imposed for the values
chosen in Table \ref{finetunetable2}, but this can always be compensated using the
rescaling (\ref{rescaling3}).

To compute observable quantities for the CMB we numerically evolve
the scalar field starting close to the saddle point, and let the
fields evolve according to the cosmological evolution equations
for non-canonically normalized scalar fields
\cite{Kallosh:2002gf, bcsq,Kallosh:2004rs,bgh}:
\begin{eqnarray}
  \label{fullequations}
    \ddot \varphi^i &+& 3H \dot \varphi^i + \Gamma^i_{jk}
    \, \dot\varphi^j \, \dot \varphi^k
    + g^{ij} \, {\partial V\over\partial\varphi^j} = 0 \,,\nonumber\\
    H^2 &\equiv & \left({\dot a\over a}\right)^2 = {8\pi G\over3} \left(\frac12
    g_{ij}\dot \varphi^i \dot \varphi^j +V\right) \,,
\end{eqnarray}
where $\varphi_i$ represent the scalar fields (${\rm Re}\, T\equiv
X$ and ${\rm Im}\, T\equiv Y$ in our case), $a$ is the scale
factor (not to be confused with the exponent in the
superpotential), and $\Gamma^i_{jk}$ are the target space Christoffel
symbols using the metric $g_{ij}$ for the set of real scalar fields $\varphi^i$ such that
${\partial^2K \over \partial
\Phi^I\partial\Phi^{J*}} \partial \Phi^I \partial \Phi^{*J}= {1\over 2} g_{ij}\partial\varphi^i\partial\varphi^{j} $.

For numerical purposes it is more convenient to write the
evolution of the fields as a function of the number $N$ of
$e$-foldings rather than time. Using
\begin{eqnarray}
a(t)\ =\ e^{N}, \qquad \qquad  \frac{d}{dt} = H \frac{d}{dN},
\end{eqnarray}
we avoid having to solve for the scale factor, instead directly
obtaining $X(N)$ and $Y(N)$.  The equations of motion are
\beqa
 X'' &=& - \left(1 - {X'^2+Y'^2\over 4 X^2}\right) \left(3X' + 2X^2{V_{,X}\over V}\right)
 + {X'^2-Y'^2\over X} \nonumber\\
 Y'' &=& - \left(1 - {X'^2+Y'^2\over 4 X^2}\right) \left(3Y' + 2X^2{V_{,Y}\over V}\right)
 + 2 {X'Y'\over X}
\eeqa
where $'$ denotes ${d\over dN}$.  The results of the numerical evolution are shown
in figure \ref{xyefold} and \ref{xypath}, using the parameters (\ref{parvalues})
and the initial conditions $X=X_{\rm saddle}$ (from (\ref{saddlepoint})) and
$Y=0.1$. Fig.\ \ref{xyefold} shows that this choice gives approximately
137 e-foldings of inflation before the fields start oscillating around one of the
local minima.  Fig.\ \ref{xypath} illustrates that the inflaton is primarily
$Y$ at the very beginning of inflation, as must be the case since $Y$ is the unstable
direction at the saddle point.  Starting at $Y=0.2$ would give 63 rather than
137 e-foldings of inflation.  However, as it has already been explained, tuning of
initial conditions is not a concern within the present model, where inflation is
topological and eternal; all possible initial conditions will be present in the
global spacetime.

\DOUBLEFIGURE[ht]{trajectories.eps, width=\hsize} {path.eps, width=\hsize}
{Evolution of $X$ (upper curve) and $Y$
(lower curve) from their initial values near the saddle point, to
one of the degenerate minima of the potential.\label{xyefold}}
{Path of inflaton trajectory in $X$-$Y$ field space.\label{xypath}}


\subsection{Experimental Constraints and Signatures}

Let us now consider the experimental constraints on and predictions of the racetrack
inflation model.  First, we must satisfy the COBE normalization on the power
spectrum of  scalar density
perturbations,  $\sqrt{P(k)} = 2\times 10^{-5}$, at the scale $k_0 \sim 10^3$ Mpc, or
equivalently at the value of $N$ which is approximately 60 e-foldings before the end
of  inflation.
We ignore isocurvature fluctuations (arising from fluctuations of
the fields orthogonal to the inflaton path shown in fig.\ \ref{xypath}) since there is
always a hierarchy between the second derivative of the potential along the path
relative to that along the orthogonal direction; then the magnitude of the scalar
power spectrum can be approximated by either
\beq
 P_1(k) = {1\over 50\pi^2}{H^4\over {\cal L}_{\rm kin}} \qquad {\rm or\ }\qquad
 P_2(k) = {1\over 150\pi^2}{V\over \epsilon}
\eeq
where the generalization of the slow-roll parameter $\epsilon$ in the two-field
case is given by
\beq
\epsilon = M_p^2{(V_{,X}\dot X + V_{,Y}\dot Y)^2\over 4\,{\cal L}_{\rm kin}\, V^2}
\eeq
We have numerically checked that the two formulas give consistent results during
the slow-roll period, and that the COBE normalization is satisfied for the
parameters given in \pref{parvalues}.

The spectral index is defined to be
\beq
n_s = 1 + {d\ln P(k)\over d\ln k} \cong 1 + {d\ln P(N)\over d N}
\eeq
where the latter approximation follows from the fact that $k= aH\cong He^{N}$ at
horizon crossing, so $d\ln k\cong dN$.  In Fig.\ \ref{pert} we plot $n_s-1$ versus
$N$, showing that $n_s\approx 0.95$ in the COBE region of the spectrum.   This red value
is typical for inflationary potentials with negative curvature. From the slope we see that $dn/d\ln k\cong -0.001$, so that the running of
the index is negligible relative to the current experimental sensitivity, also typical
for models in which the flatness of the potential is not punctuated by any special
features.

The value of the spectral index $n_s\approx 0.95$ and the smallness of the running index appear to be pretty stable with respect to various modifications of the model; we were unable  to alter these results by changing various parameters. This makes racetrack inflation testable; at present, the best constraint on $n_{s}$ is $n_{s} = 0.98 \pm 0.02$ \cite{Seljak:2004xh}, which is compatible with our results. Future experiments will be able either confirm our model or rule it out; in particular, Planck satellite will measure $n_{s}$ with accuracy better than $0.01$.

\FIGURE [h!]{ \centering\leavevmode\epsfysize=6cm \epsfbox{pert.eps}
\caption[fig10] {Deviation of the scalar power spectral index from 1 ($n-1$)
versus $N$ (equivalent to $n-1$ versus $\ln k$) for the typical racetrack
model.} \label{pert} }

It is interesting
that experimental tests will be able to discriminate for or against the model in
the not-too-distant future.  The presence of a tensor contribution will not provide
any test in the example we have considered since the scale of inflation is
$V^{1/4} \sim 10^{14}$ GeV, far below the $3\times 10^{16}$ GeV threshold needed for
producing observable gravity waves.

In order to formulate the theory of reheating in our scenario one needs to find
a proper way to incorporate the standard model matter fields in our model. In
the KKLT scenario there are two possible places where the standard model
particles can live: on (anti) D3 branes at the end of the throat \cite{cgqu} or
on the wrapped D7 branes, after some twisting and/or turning-on  of magnetic
fields. If the standard model fields live on a D7 brane, the axion couples to
the  vector fields like $Y\, F_{\mu\nu}  \tilde F_{\mu\nu}$ and the volume
modulus as $ X F_{\mu\nu}  F_{\mu\nu}$. To study the reheating one has to find
the decay rate of the $X$ and $Y$ fields to the vector particles and the
corresponding reheating temperature. A preliminary investigation of this
question along the lines of Ref. \cite{KLS97} shows that reheating in this
scenario can be rather efficient.\footnote{This situation will be the same in
heterotic string realizations of our scenario (with the role of the $T$ field
taken by the dilaton $S$) in which the dilaton field couples universally to
hidden and observable gauge fields.} Explicit models of this type are yet to be
constructed. If the standard model is on D3 branes, we would have to consider
other couplings of the $T$-field to matter fields.  We hope to return to the
theory of reheating in the racetrack inflation scenario in a separate
publication.

\section{The overshooting problem and initial conditions for the racetrack inflation}

Even though we already discussed advantages of eternal topological inflation
from the point of view of the problem of initial conditions, we will revisit
this issue here again, emphasizing specific features of eternal inflation in
the context of string theory.

There is a well-known problem related to initial conditions in string cosmology
\cite{Brustein:nk}. This problem, in application to the KKLT-based models, can
be formulated as follows. Even though there is a KKLT minimum of the potential
with respect to the dilaton field and the volume modulus $X$, this minimum is
separated from the global Minkowski minimum at $X \to \infty$ by a relatively
small barrier. The height of the barrier depends on the parameters of the
model, but in the simplest models considered in the literature it is 10 to 20
orders of magnitude smaller than the Planck density. In particular, for the
parameters of our model, the height of the barrier is $2 \times 10^{-16}$ in
Planckian units, see Fig. 1. Generically, one could expect that soon after the
big bang, the energy density of all fields, including the field $X$, was many
orders of magnitude greater than the height of the barrier. If, for example,
the field $X$ initially was very small, with energy density much larger than $2
\times 10^{-16}$, then it would fall down along the exponentially steep
potential, easily overshoot the KKLT minimum, roll over the barrier, and
continue rapidly rolling towards $X \to \infty$. This would correspond to a
rapid decompactification of the 4D space.

There are several ways to avoid this problem. One may try, for example, to
evaluate the possibility that instead of being born in a state with small $X =
{\rm Re}~T$, the universe was created ``from nothing'' in a state corresponding
to the inflationary saddle point of the potential $V(T)$. This would resolve
the problem of overshooting. However, at the first glance, instead of solving
the problem of initial conditions, this only leads to its even sharper formulation.

Indeed, according to \cite{Linde:1983mx,Vilenkin:1984wp}, the probability of
quantum creation of a closed universe in a state corresponding to an extremum
of its effective potential is given by $P \sim \exp\left(-{24\pi^{2}\over V(T)}\right)$,
where the energy density is expressed in units of the Planck density. The same
expression appears for quantum creation of an open universe
\cite{Linde:1998gs}. This leads to an alternative formulation of the problem of
initial conditions in our scenario: The probability that a closed (or open)
universe is created from ``nothing'' at the saddle point with $V \sim 2\times
10^{{-16}}$ is exponentially suppressed by the factor of $ P \sim
\exp\left(-{10^{18}}\right)$. This result, taken at its face value, may look
pretty upsetting.  (The simplest models of chaotic inflation, which can begin
at $V(\phi) = O(1)$, do not suffer from this problem \cite{book}.)

In Section \ref{topinfl} we outlined a possible resolution of this problem:
Even if the probability of proper initial conditions for eternal topological
inflation is extremely small, the parts of the universe where these conditions
are satisfied enter the regime of eternal inflation, producing infinite amount
of homogeneous space where life of our type is possible.  Thus, even if the
fraction of the universes (or of the parts of our universe) with inflationary
initial conditions is exponentially suppressed, one may argue that eventually
most of the observers will live in the parts of the universe produced by
eternal topological inflation.

Here we would like to strengthen this argument even further, by finding the
conditions which may allow us to remove the  exponential suppression of the
probability of initial conditions for the low energy scale inflation. In order
to do it, let us identify the root of the problem of the exponential
suppression: At the classical level,  the minimal size of a closed de Sitter
space is $H^{-1}$. Quantum creation of a closed inflationary universe is
described by the tunneling from the state with the scale factor $a = 0$ (no
universe) to the state where the size of the universe becomes equal to its
minimal value $a = H^{-1}(T)$. Exponential suppression appears because of the
large absolute value of the Euclidean action on the tunneling trajectory
\cite{Linde:1983mx,Vilenkin:1984wp}. One could expect that for an open universe
there is no need for tunneling because there is no barrier for the classical
evolution of an open universe from $a =0$ to larger $a$. However, an instant
creation of an {\it infinite} homogeneous open universe is problematic
(homogeneity and horizon problems). The only known way to describe it is to use
a different analytic continuation of the same instantons that were used for the
description of quantum creation of a closed universe \cite{Hawking:1998bn},
which again leads to the exponential suppression of the probability
\cite{Linde:1998gs}.

Fortunately, there is a simple way to overcome this problem
\cite{Linde:2003hc}, which is directly related to the standard Kaluza-Klein
picture of compactification in string theory. In this picture, it is natural to
assume that {\it all} spatial dimensions enter the theory democratically, i.e.
all of them are compact, not necessarily because of the curvature of space (as
in the closed universe case), but because of its nontrivial topology.  Then
inflation makes the size of 3 of these dimensions exponentially large, whereas
the size of 6 other dimensions remains fixed, e.g., by the KKLT mechanism.

For example, one may consider a toroidal compactification of a flat universe,
or compactification of an open universe. This is a completely legitimate
possibility, which was investigated by many authors; see e.g.
\cite{topol1,topol2,topolbf,topol3}. It does not contradict any observational
data if inflation is long enough to make the size of the universe much greater
than $10^{10}$ light years.

An important feature of a topologically nontrivial compact flat or open dS
space is that (ignoring the Casimir effect which is suppressed by
supersymmetry) its classical evolution can continuously proceed directly from
the state with $a(t) \to 0$, without any need of tunneling, unlike in the
closed universe case \cite{Zeldovich:1984vk}. As a result, {\it there is  no
exponential suppression of the probability of quantum creation of a compact
flat or open inflationary universe with $V(T)\ll 1$} \cite{Linde:2003hc}. This
observation removes the main objection against the possibility of the low
energy scale inflation starting at an extremum of the effective potential.

Moreover, in an eternally inflating universe consisting of many de Sitter parts
corresponding to the string theory landscape, the whole issue of initial
conditions should be formulated in a different way \cite{LLM,Linde:2004kg}. In
such a universe, inflation is always eternal because of the incomplete decay of
metastable dS space, as in old inflation. Therefore evolution of such a
universe will produce infinitely large number of exponentially large parts of
the universe with different properties \cite{LLM,susskind}. Even if in many of
such parts space will become 10D after the scalar field $T$ overshoots the KKLT
barrier, this will be completely irrelevant for the evolution of other parts of
the universe where space remains 4D. In some parts of 4D space there was no
inflation, and adiabatic perturbations with a flat spectrum could not be
produced. Even though such parts initially may be quite abundant, later on they
do not experience an additional inflationary growth of their volume.
Observations tell us that we do not live in one of such parts. Meanwhile
inflationary trajectories starting at the saddle point describe eternally
inflating parts of the universe which produce an indefinitely large volume of
homogeneous 4D space where observers of our type can live. This observation
goes long way towards resolving the problem of initial conditions in our
scenario.

\section{Irrational Racetrack Inflation and the Cosmological Constant}

In order to achieve a successful cosmological scenario, one needs
to make at least two fine-tunings. First of all, we must fine-tune
the uplifting of the AdS potential in the KKLT scenario to obtain
the present value of the cosmological constant $\Lambda \sim
10^{-120}$ in units of Planck density. Then we must fine-tune the
parameters of our model in order to achieve a slow-roll inflation
with the amplitude of density perturbations ${\delta\rho\over
\rho} \sim 10^{-5}$.

One way to approach the fine-tuning problem is to say that the
smallness of the cosmological constant, as well as the stage of
inflation making our universe large and producing small density
perturbations are necessary for the existence of life as we know
it. This argument may make sense in the context of the eternal
inflation scenario, but only if there are sufficiently many vacuum
states with different values of the cosmological constant, and one
can roll down to these states along many different inflationary
trajectories.

In this respect, the possibility that the eternally inflating
universe becomes divided into many exponentially large regions
with different properties \cite{book, anthropic} and the related
idea of the string theory landscape describing enormously large
number of different vacua may be very helpful
\cite{bp,susskind,douglas}. However, even in this case one must
check whether the set of parameters which are possible in string
theory is dense enough to describe theories with $\Lambda \sim
10^{-120}$ and ${\delta\rho\over \rho} \sim 10^{-5}$.

\begin{figure}[h!]
\centering\leavevmode\epsfysize=7cm \epsfbox{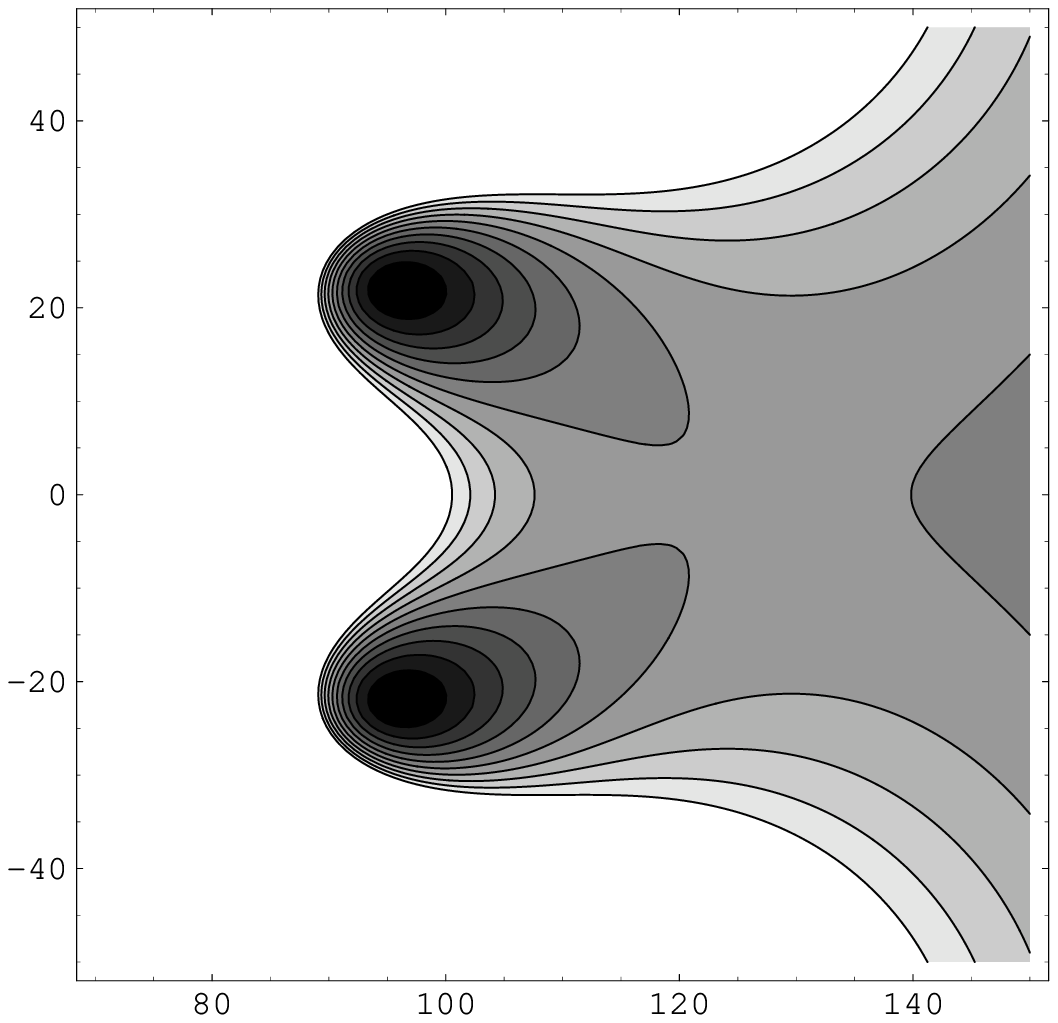} \hskip 0.7cm
\epsfysize=7cm \epsfbox{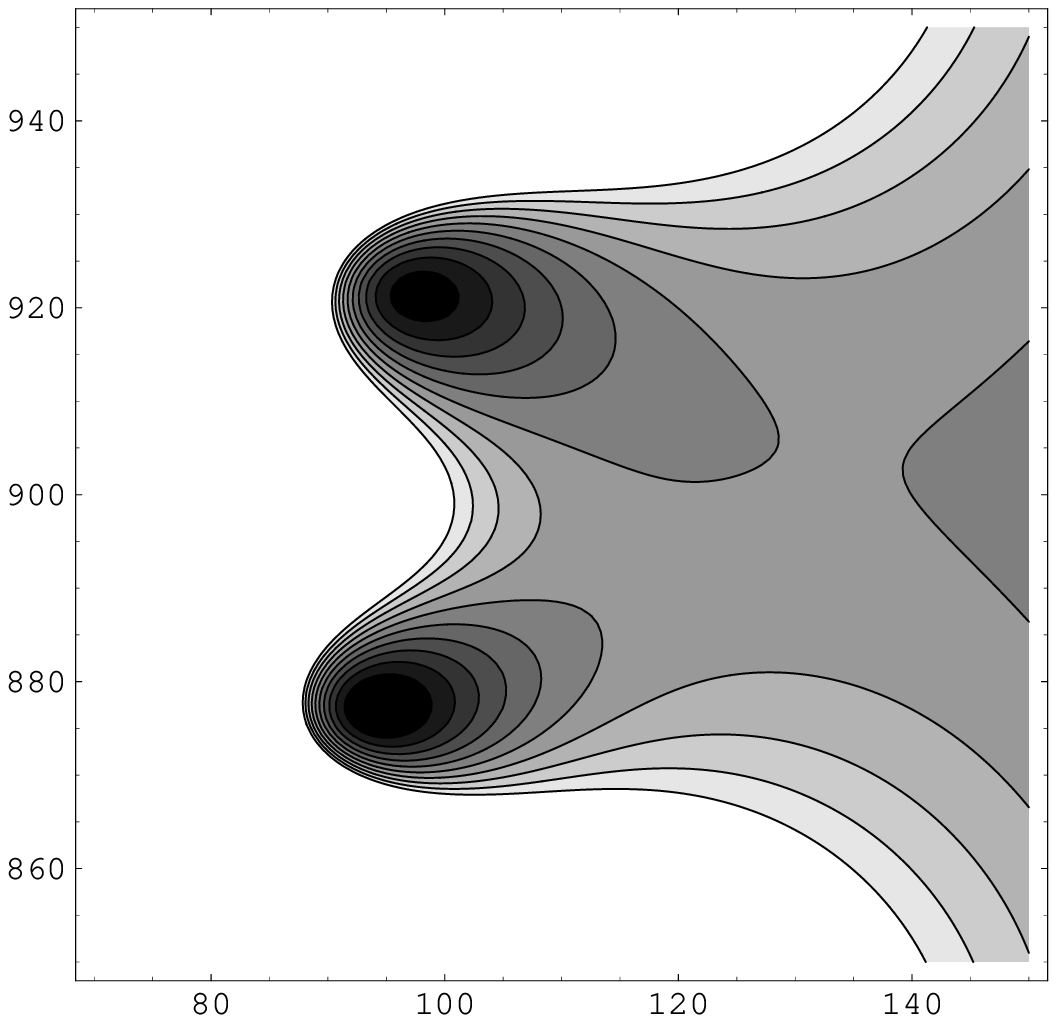}

\caption[fig2] {Contour plot of the effective potential in the vicinity of the
saddle point at $Y =0$ (left panel) and at $Y \sim 900$  (right
panel).\label{two}}
\end{figure}

One way towards making the choice of various vacua infinitely rich
is related to the irrational axion scenario \cite{Banks:1991mb}.
In our previous discussion of the effective potential of our
model, Eq. (\ref{potential1}), we assumed that $a = {2\pi\over M}$
and $b ={2\pi\over N}$, where $N$ and $M$ are integers.
In this
case the potential is periodic in $Y$, with a period $MN$, or less,
 $MN/(M-N)$ if $M/(M-N)$ and $N/(M-N)$ are integers. For
example, if one takes $M = 100$ and $N = 90$, the potential will
have identical inflationary saddle points for all $Y = n(MN/(M-N)) =
900n$,  where $n$ is an arbitrary integer.

However, if one assumes that instead of being an integer, at least
one of the numbers $M$, $N$ is irrational, then in the infinitely
large  interval of all possible values of the axion field $Y$
values one would always find vacua with all values of $\Lambda$
varying on a very large scale comparable with the height of the
potential barrier in the KKLT potential. Similarly, if we, for
example, start with an irrational number $M$ very close to $1000$,
then at $Y=0$ we will have the same scenario as in the model
described above, but at large $Y$ we will have all kinds of
extrema and saddle points, leading to various stages of inflation
with different amplitude of density perturbations.

To illustrate this idea, we give here the contour plots of two
subsequent saddle points which appear in our theory under a very
mild modification: We replace $M = 100$ by $M = 100+{\pi\over
100}$. It immediately produces an infinitely large variety of
different shape saddle points and minima of different depth; we show two
different saddle points separated by $Y \approx 900$ in Fig.
\ref{two}.  The shape and the depth of the minima of these
potentials can be changed  in an almost continuous way. This
reflects  the situation with irrational winding line on the torus
for incommensurate frequencies:  the system will never return to
its original position, it will cover the torus densely, coming
arbitrary close to every point.

The main question is whether one can find a version of string
theory with the superpotential involving irrational $M$. Whereas
no explicit examples of such models are known at present, the idea
is so attractive that we allow ourselves to speculate about
its possible realizations. One can try to use the setting of
 \cite{Martinec:2001hh} where the non-commutative solitons on
 orbifolds
are studied and some relation to irrational axions is pointed out.
 One can also look for the special contribution to the Chern-Simons
 term on
D7 due to the 2-form fluxes. The standard $a FF^*$ axion coupling
originates
 from the term $\int _{M^9} d[C e^{{\cal F}}]$ where one
finds $d[C e^{\cal F}] \sim da\wedge F\wedge F\wedge J\wedge J$ and
$J$ is
the K\"ahler form and $da\wedge J\wedge J$ comes from the self-dual
five-form.
 Perhaps it is not impossible  to find in addition to this term a
 contribution
 where one or both of the K\"ahler forms are replaced by the 2-form
 fluxes
${\cal F}= F-B$. Here the 2-form $B$ is not quantized and is related
 to the
 non-commutativity parameter in the internal space. This may lead to
 some
 new coupling of the form $ q a FF^*$ where $q$ is irrational as
 proposed
 in \cite{Banks:1991mb}.

\section{Conclusions}

We present what we believe is the simplest model for inflation in
a string-theoretic scenario. It requires a certain amount of
fine-tuning
 of the parameters of the non-perturbative racetrack superpotential.
Inflation only occurs due to the
existence of a nontrivial potential lifting the complex K\"ahler
structure modulus whose real part is the overall size of the
compact space and the imaginary part coming from the type IIB
four-form is the inflaton field. The periodicity properties in
this direction make the scalar potential very rich and we find
explicit configurations realizing the slow-roll/topological
inflation setting that give rise to eternal inflation and density
perturbations in our patch consistent with the COBE normalization
and spectral index inside (but close to the edge of) the current observational bounds.
We find this to be very encouraging.

It is appropriate to compare our scenario with the other direction
for obtaining inflation in string theory, namely the brane/brane
\cite{dvalitye,d3d7} or brane/antibrane \cite{bmnqrz,dss} inflationary
scenarios in which the inflaton field corresponds to the
separation of the branes $\psi$ and inflation ends by the
appearance of an open string tachyon in a stringy realization of
hybrid inflation \cite{bmnqrz,d3d7}.  Clearly our racetrack scenario is
much simpler since there is no need to introduce the brane
configurations with their separation field $\psi$ to obtain
inflation. The analysis of \cite{kklmmt} pointed out the
difficulty in obtaining brane inflation once a mechanism for
moduli fixing is included. Various  extensions and improvements of the
basic
 scenario were considered in \cite{kklmmt,bcsq,shamit}, they typically
 require
 a fine-tuning. At present only  D3/D7 brane inflation model
 compactified
 on $K3\times {T_2\over Z_2}$ with volume stabilization
does not seem to require a fine-tuning \cite{KKL}.

In addition,  in brane inflation scenario the tachyon
potential can give rise to topological defects such as cosmic
strings \cite{bmnqrz,kklmmt,henry2} which may have important observational
implications \cite{henry2,cosmicstrings}.
Furthermore, having an implementation of inflation in string
theory, independent of brane inflation, diminishes in some sense
the possible observational relevance of cosmic strings as being a
generic implication of the brane inflation scenario. In racetrack
inflation, cosmic strings need not be generated after inflation.

On the other hand there is no inconsistency to combine both
scenarios. For instance, in realistic generalizations of the KKLT
scenario including standard model branes \cite{cgqu}, the throat
where the standard model lives should be significantly warped,
whereas brane/antibrane inflation required a mild warping in the
inflation throat in order to be consistent with the COBE
normalization \cite{bcsq}. Also the fine tuning involved suggested
two or more stages of inflation with some 20-30 e-foldings at a
high scale and the rest at a small scale, probably close to 1 TeV.
This, besides having interesting observational implications, may
also help to the solution of the cosmological moduli problem
\cite{modprob} which usually needs a late period of inflation. We
could then foresee a scenario in which racetrack inflation is at
work at high scales and the low-scale inflation is provided by
brane/antibrane collision in the standard model throat.
See \cite{graham} for a previous discussion of low-scale inflation.

There are many avenues along which to generalize our approach.
Modular inflation could also be obtained by using more general
modular superpotentials than those of the racetrack type
considered here. Also, potentials for many moduli fields are only
starting to be explored and some of them have the advantage that a
nonvanishing superpotential has been obtained \cite{douglas}. It
would also be interesting to investigate if the irrational cases
proposed here are actually obtainable from string theory. We hope
to report on some of these issues in the future.

\section{Acknowledgements} We would like to thank R.~Brustein, E.~Copeland,  S.~P.~de Alwis,  G.~
Dvali, A. Font, S. Gratton,
 A.\
Gruzinov, A.\ Iglesias, X.\ de la Ossa, L. Ib\'a\~nez, S. Kachru, N.\  Kaloper,
G.\  Ross,  S.\  Sarkar, R.\ Scoccimarro, E.\
Silverstein, H. Tye  and A.\  Vilenkin for useful conversations.
We also thank the organizers of the `New
Horizons in String Cosmology' workshop at the Banff International
Research Station for providing the perfect environment for
completing this work and for contemplating the landscape.  C.B. and
J.C. are funded by grants from NSERC (Canada), NATEQ (Qu\'ebec) and
McGill University. The work of M.G.-R.\  is
supported by D.O.E. under grant DE-FG02-92ER40706.  J.J.B-P. is supported by
the James Arthur Fellowship at NYU. C.E. is partially funded by EPSRC.
The work of R.K. and A.L. is supported
by the NSF (US) grant 0244728. F.Q. is
partially funded by PPARC and a Royal Society Wolfson award.

\end{document}